\documentclass[letterpaper]{article} 
\usepackage{aaai25}  
\usepackage{times}  
\usepackage{helvet}  
\usepackage{courier}  
\usepackage[hyphens]{url}  
\usepackage{graphicx} 
\usepackage{natbib}  
\usepackage{caption} 
\usepackage{booktabs}  
\usepackage{microtype} 
\usepackage{multirow}
\usepackage{amsmath}
\newcommand{\mypara}[1]{{\vspace{1.5mm}\noindent\textbf{#1}}}
\usepackage{algorithm}
\usepackage{algorithmic}
\usepackage{url}

\usepackage{newfloat}
\usepackage{listings}
\DeclareCaptionStyle{ruled}{labelfont=normalfont,labelsep=colon,strut=off} 
\floatstyle{ruled}
\newfloat{listing}{tb}{lst}{}
\floatname{listing}{Listing}
\title{SECodec: Structural Entropy-based Compressive Speech Representation Codec for Speech Language Models}
\author{Linqin Wang\textsuperscript{1,2}, Yaping Liu\textsuperscript{1,2}, Zhengtao Yu\textsuperscript{1,2} \thanks{ \quad Corresponding author},  
Shengxiang Gao\textsuperscript{1,2},  Cunli Mao\textsuperscript{1,2},  \\ Yuxin Huang\textsuperscript{1,2}, Wenjun Wang\textsuperscript{1,2}, Ling Dong\textsuperscript{1,2}
\\ \textsuperscript{1} Faculty of Information Engineering and Automation, \\ Kunming University of  Science and Technology, Kunming, China \\  \textsuperscript{2} Yunnan Key Laboratory of Artificial Intelligence, Kunming, China \\ 
  }
\affiliations{

    \textsuperscript{1} Faculty of Information Engineering and Automation, Kunming University of  Science and Technology, Kunming, China \\  \textsuperscript{2} Yunnan Key Laboratory of Artificial Intelligence, Kunming, China \\
    \texttt{\{linqinwang7767\}@163.com},  \texttt{\{ztyu,gaoshengxiang.yn\}@hotmail.com},\\
  \texttt{\{maocunli,lyp845935161,huangyuxin2004\}@163.com}, \texttt{20203104003@kust.edu.cn}, \texttt{ling.dong@kust.edu.cn}
}

\usepackage{bibentry}

\begin{document}

\maketitle

\begin{abstract}
With the rapid advancement of large language models (LLMs), discrete speech representations have become crucial for integrating speech into LLMs. Existing methods for speech representation discretization rely on a predefined codebook size and Euclidean distance-based quantization. However, 1) the size of codebook is a critical parameter that affects both codec performance and downstream task training efficiency. 2) The Euclidean distance-based quantization may lead to audio distortion when the size of the codebook is controlled within a reasonable range.
In fact, in the field of information compression, structural information and entropy guidance are crucial, but previous methods have largely overlooked these factors.
Therefore, we address the above issues from an information-theoretic perspective, we present SECodec, a novel speech representation codec based on structural entropy (SE) for building speech language models. 
Specifically, we first model speech as a graph, clustering the speech features nodes within the graph and extracting the corresponding codebook by hierarchically and disentangledly minimizing 2D SE. Then, to address the issue of audio distortion, we propose a new quantization method. This method still adheres to the 2D SE minimization principle, adaptively selecting the most suitable token corresponding to the cluster for each incoming original speech node. 
Furthermore, we develop a Structural Entropy-based Speech Language Model (SESLM) that leverages SECodec. 
Experimental results demonstrate that SECodec performs comparably to EnCodec in speech reconstruction, and SESLM surpasses VALL-E in zero-shot text-to-speech tasks. 
Code, demo speeches, speech feature graph, SE codebook, and models are available at \url{https://github.com/wlq2019/SECodec}.

\end{abstract}

%

\section{Introduction
}
Large language models (LLMs)~\cite{achiam2023gpt,touvron2023llama} have exhibited exceptional capabilities in a wide range of natural language processing tasks. This success has spurred extensive research efforts in developing speech language models~\cite{zhang2023speechtokenizer,huang2023repcodec,borsos2023audiolm}, leading to notable advancements in numerous speech processing applications~\cite{wang2023neural,tu2024smart,rubenstein2023audiopalm,dong2023polyvoice,tu2023view}. To bridge the gap between continuous speech and token-based language models, a crucial method called speech discretization is employed. This process transforms an audio signal into a finite set of tokens. By converting speech into discrete tokens, language models are able to predict future semantic content and generate coherent and realistic speech with long-term consistency~\cite{nguyen2022discrete,tu20222}.

Current discrete speech representations for
speech language models can be categorized into three types: semantic
tokens, acoustic tokens, and hybrid/unified tokens~\cite{borsos2023audiolm,zhang2023speechtokenizer}. 1) Semantic tokens~\cite{hsu2021hubert,baevski2020wav2vec,chung2021w2v} are typically generated from self-supervised pre-trained models using masked language modeling as the training objective, which are derived through $k$-means clustering on representations from a specific intermediate layer, resulting in sequences with a one-dimensional structure. Speech language models that use semantic tokens~\cite{lakhotia2021generative,zhang2023speechgpt,hassid2024textually} can be externally connected to a vocoder for speech synthesis. While these models effectively capture semantically accurate content, the resulting speech generation often suffers from poor quality and a loss of acoustic details.
2) Acoustic tokens~\cite{zeghidour2021soundstream,defossez2022high,yang2023hifi,du2024funcodec} are extracted from neural audio codecs, which use reconstruction as the training objective. By employing residual vector quantization (RVQ)~\cite{gray1984vector,vasuki2006review} with hierarchical quantizers for discretization, acoustic tokens are represented as matrices with two dimensions: timesteps and quantizers. 
VALL-E~\cite{wang2023neural} is a representative model of speech language models that utilize acoustic tokens. Despite achieving impressive zero-shot text-to-speech (TTS) capabilities, it still faces issues such as inaccurate content, stemming from the complex information contained within acoustic tokens. 3) 
Hybrid or unified tokens~\cite{borsos2023audiolm,zhang2023speechtokenizer} employ different strategies to combine semantic tokens and acoustic tokens. Hybrid tokens adopt a hierarchical approach, encompassing both semantic token language models and acoustic token language models, to capture content information and acoustic details, respectively~\cite{dong2023polyvoice,borsos2023audiolm,rubenstein2023audiopalm}. Recently, unified tokens, exemplified by SpeechTokenizer~\cite{zhang2023speechtokenizer}, have emerged. These tokens distill semantic information into acoustic tokens, effectively unifying semantic and acoustic representations. SpeechTokenizer has achieved superior results in downstream tasks such as speech synthesis. 
The ideal speech representation for speech language models should meet two key characteristics: i) Effective preservation of speech information; ii) Sufficient compressiveness for efficient training of speech language models.
However, i) existing speech discretization methods rely on $k$-means to initialize the codebook space, the size of codebook is a critical parameter that significantly impacts the performance of the codec and the training efficiency of downstream tasks, yet its size is typically determined through empirical judgment. ii) Additionally, when attempting to control the size of the codebook within a reasonable range, the quantization process, which relies on Euclidean distance, may lead to substantial differences between codebook's vector and original vector, resulting in audio distortion. These issues result in a loss of information and produce overly long tokens that are difficult to train, thereby impairing overall performance.

In this work, we address the aforementioned issues from an information-theoretic perspective, drawing inspiration from structural entropy (SE)~\cite{li2016structural,cao2024hierarchical}, a metric that assesses the amount of information contained in a graph~\cite{yang2024hierarchical,zeng2024scalable,yang2024adaptive,yang2024sebot,cao2024multi,peng2024unsupervised,zou2024multispans}. We present SECodec, a novel speech representation codec tokenizer based on structural entropy, which can automatically determine the appropriate codebook size and integrate structural information into the quantization process. 
Experiments demonstrate that these approaches effectively mitigate the information loss problem prevalent in existing speech discretization methods.
Our main contributions are:
\begin{itemize}
    \item  We model the speech representation codec from an information-theoretic perspective. Compared to previous methods that use $k$-means, the proposed SECodec, by introducing structural information and entropy guidance, learns a more compressive and informative codebook without requiring a predetermined codebook size. To the best of our knowledge, we are the first to apply structural entropy (SE) minimization for a speech representation codec.
    \item To address the issue of audio distortion when controlling codebook size, we propose a new quantization method that iteratively selects appropriate clusters for the added original speech features using a SE heuristic function. This approach significantly enhances the quality of information in the speech tokens.
    \item Extensive experimental results demonstrate that SECodec performs comparably to EnCodec in speech reconstruction, while SESLM surpasses VALL-E in zero-shot text-to-speech tasks on a multi-speaker benchmark dataset. 
\end{itemize}

\begin{figure*}[!t]
\vspace{-1.0em}
\centering
\includegraphics[width=6in]{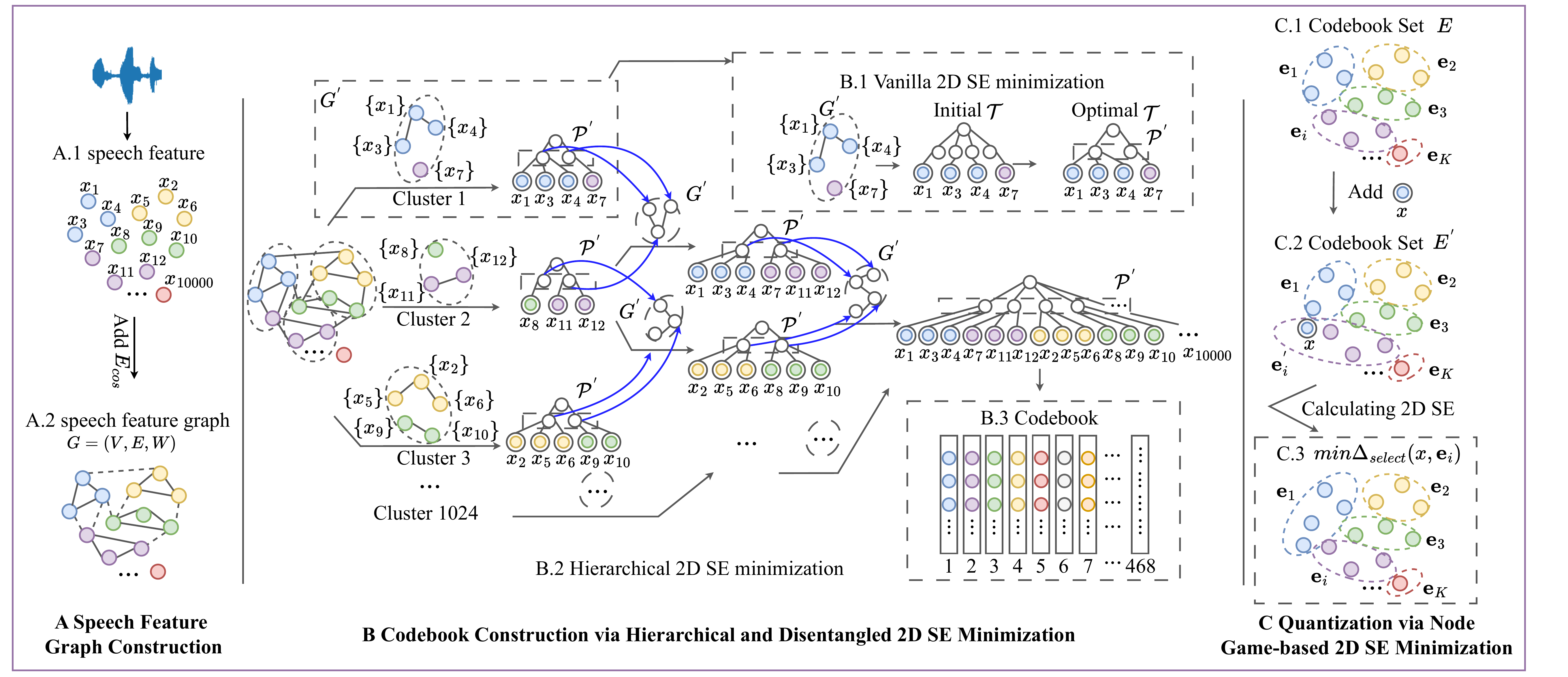}
\caption{Framework of proposed model.}
\label{model}
\vspace{-1.0em}
\end{figure*}
\section{Preliminary
}
Structural entropy (SE)~\cite{li2016structural} is defined as the minimum number of bits to encode the vertex that is accessible with a step of random walk on a graph. SE is a measurement of graph
complexity by encoding tree structures via characterizing
the uncertainty of the hierarchical topology of graphs. The structural entropy of graph $G$ is defined on an associated
encoding tree $\mathcal{T}$, revealing the amount of uncertainty that
remained in $G$ after encoded by $\mathcal{T}$. Through structural
entropy minimization, the optimized hierarchical clustering
result of vertices in $G$ is retained by $\mathcal{T}$. We present the formal definitions of encoding tree and SE as follows.
 
\mypara{Definition 1. }Let $G=(V,E,W)$ be an undirected weighted graph, where $V=\{v_1,...,v_n\}$ is the vertex set, $E$ is the edge set, and $W \in R^{n \times n}$ is the edge weight matrix.

1) \textit{The encoding tree $\mathcal{T}$ of $G$ is a hierarchical rooted tree where each tree
node $\alpha$ associates with a vertex set $T_{\alpha}$.}

2) \textit{The root node $\lambda$ of
$\mathcal{T}$ associates with $T_{\lambda}=V$ and each leaf node $v$ associates with $T_v$ containing a vertex in $V$.}

3) \textit{For each non-leaf
node $\alpha \in \mathcal{T}$, the successors of $\alpha$ are associated with disjoint vertex subsets, and the union of these subsets is $T_{\alpha}$.}

\mypara{Definition 2. }The structural entropy of $G$ given by $\mathcal{T}$ is defined as follows:
\begin{equation}
\begin{split}
\mathcal{H}^{\mathcal{T}}(G)=\sum_{\alpha \in \mathcal{T},\alpha \neq \lambda}\mathcal{H}^{\mathcal{T}}(G;\alpha)=\\ \sum_{\alpha \in \mathcal{T},\alpha \neq \lambda}-\frac{g_{\alpha}}{\mathcal{V}_G}log_2\frac{\mathcal{V}_{\alpha}}{\mathcal{V}_{{\alpha}^-}},
\end{split}
\end{equation}
where $\mathcal{H}^{\mathcal{T}}(G;\alpha)$ is the assigned structural entropy of $\alpha$, $g_{\alpha}$
is the cut, i.e., the sum of edge weights between vertices in and not in $T_{\alpha}$, $\mathcal{V}_{\alpha}$ and $\mathcal{V}_{G}$ are the volumes, i.e., the sum of vertex degrees in $T_{\alpha}$ and $G$, respectively. The structural
entropy of $G$ is defined as $\mathcal{H}(G)=\mathop{\min}_{\mathcal{T}} \{ \mathcal{H}^{\mathcal{T}}(G) \},$
where $\mathcal{T}$ ranges over all possible encoding trees. The vertex sets associated with tree nodes form a clustering
of vertices in $V$.

\mypara{Definition 3. }The $K$-D structural entropy is the structural entropy given by the encoding trees with the height of at most $K$. When $K=2$, the encoding
tree represents graph partitioning, which can be used to perform partitioning clustering. A $2$-D encoding tree $\mathcal{T}$ can
be formulated as a graph partitioning $\mathcal{P}=\{ \mathcal{X}_1,\mathcal{X}_2,...,\mathcal{X}_L\}$ of $V$ , where $\mathcal{X}_i$ is a vertex subset called module associated with the $i$-th children of root $\lambda$. The structural entropy of $G$
given by $\mathcal{P}$ is defined as:
\begin{equation}
\begin{split}
\mathcal{H}^{\mathcal{P}}(G)=-\sum_{\mathcal{X} \in \mathcal{P}}\sum_{v_i \in \mathcal{X}}\frac{g_i}{\mathcal{V}_G}log_2\frac{d_i}{\mathcal{V}_\mathcal{X}}\\ -\sum_{\mathcal{X} \in \mathcal{P}}\frac{g_\mathcal{X}}{\mathcal{V}_G}log_2\frac{\mathcal{V}_\mathcal{X}}{\mathcal{V}_G},
\end{split}
\end{equation}
where $d_i$ is the degree of vertex $v_i$, $g_i$ is the cut, i.e., the sum
of edge weights connecting $v_i$ and other vertices, $\mathcal{V}_\mathcal{X}$ and $\mathcal{V}_G$ are the volumes, i.e., the sum of vertex degrees in module
$\mathcal{X}$ and graph $G$, respectively, and $g_\mathcal{X}$ is the cut, i.e., the sum
of edge weights between vertices in and not in module $\mathcal{X}$.

\section{The Method
}
Figure~\ref{model} presents an overview of SECodec. Our model is based on the RVQ-GANs~\cite{du2024funcodec} framework, akin to SoundStream~\cite{zeghidour2021soundstream} and EnCodec~\cite{defossez2022high}. However, we employ 2D structural entropy to optimize both the codebook initialization and the quantization process, resulting in more compressive codebook and more informative tokens.
We begin by formalizing the task. Subsequently, we propose a novel structural entropy-based approach for codebook construction. We then present our informative quantization process. Finally, we introduce the training objective and design the SESLM.

\subsection{Problem Formalization}
Considering the input speech feature $X = [x_1,...,x_T ] \in R^{H \times T}$ from a pre-trained convolutional network, where $H$ is the dimension of the speech representation and $T$ is the length of the sequence. we construct a speech feature
graph $G = (V,E,W)$. Here $V = \{v_1, v_2, ..., v_n\}$ is the
set of vertices corresponding to speech features in $X$ , $E$ represents the set of edges connecting the vertices, and $W$ represents the set of edge weights measuring the similarities between every frame of speech feature. For
two frames of speech feature $x_i,x_j \in X$, we measure their the cosine similarity.  Partitioning $G$ results in $\{\textbf{e}_1,...,\textbf{e}_i,...,\textbf{e}_j,...,\textbf{e}_K\}, \textbf{e}_i \subset
V, \textbf{e}_i \cap \textbf{e}_j = \emptyset$, which represents a partition of $V$ containing $K$ clusters (sets) of speech features. These clusters correspond to the codebook $\mathcal{E}_{codebook}= [\textbf{e}_1,...,\textbf{e}_K]$.

\subsection{Codebook Construction via Hierarchical and Disentangled 2D SE
Minimization}
Speech feature graph partitioning decodes $G$ into $\mathcal{P}$, which defines the size of the codebook in the form of speech feature clusters. A faithful decoding of the speech feature correlations in $G$ assigns related speech features to the same cluster and unrelated ones to different clusters. Previous RVQ-based speech codec methods use $k$-means to initialize the codebook space. These empirically defined codebooks, which must be predetermined, lead to a loss of information and result in overly long tokens that are difficult to train, consequently impairing overall performance. To address this issue, SECodec conducts codebook partitioning under the guidance of 2D structural entropy (SE) minimization. This approach reveals the essential second-order (cluster-wise) structure inherent in the raw graph without prior knowledge of the number of speech feature clusters.

\citeauthor{li2016structural} (\citeyear{li2016structural}) propose a vanilla greedy 2D structural entropy (SE) minimization algorithm that repeatedly merges any two nodes in the encoding tree $\mathcal{T}$ resulting in the largest decrease in 2D SE until reaching the minimum possible value. This process partitions a graph without supervision or a predetermined total number of clusters. However, this vanilla 2D SE minimization algorithm has a time complexity of $O(|V|^3)$, making it prohibitively slow for large and complex graphs. 
Furthermore, the ultimate goal of our clustering is to construct codebook. The column vectors in the codebook need to be spatially distributed as far apart as possible and avoid overlapping to ensure effective representation and diversity of the speech features. To address these challenges, we propose to minimize 2D
SE for construct codebook in a hierarchical and disentangled manner, shown in Algorithm~\ref{algorithm:codebook}. Specifically, each speech feature  $x_1,...,x_T$  is placed in its own cluster (line 1). These clusters are then divided into subsets of size $n$ (line 3-5), and within each subset, the vanilla greedy algorithm is used to merge the clusters into new ones (lines 6-16). The newly formed clusters proceed to the next iteration (line 17). This iterative process continues until all speech feature clusters are considered simultaneously (lines 18-19). If no clusters within a subset can be merged at any point, the subset size $n$ is increased to allow more clusters to be considered together for potential merging (lines 20-21). Finally, we extract the corresponding codebook by minimizing the mutual information between each vector within the codebook (lines 22-34). Figure~\ref{model}A shows the speech feature graph construction on nodes $x_1$ to $x_{10000}$. Figure~\ref{model}B illustrates codebook construction process: initially $x_1$ to $x_{10000}$ are in separate clusters. Clusters of size $n = 1024$ are considered at a time to form a subgraph $G'$. Clusters in each $G'$ are merged using the vanilla 2D SE minimization to form $\mathcal{P}'$ (Figure ~\ref{model}B.1). The partitions from the previous iteration are carried over to the next, as shown by the blue curved arrows in Figure~\ref{model}B.2. The process concludes when a $\mathcal{P}'$ that encompasses all the speech features is achieved. To further enhance the codebooks' ability to represent the diversity of speech, we introduce a mutual information learning algorithm to disentangle the central features of each cluster in $\mathcal{P}'$. 
For the disentanglement between $\textbf{e}_i$ and $\textbf{e}_j$. The variational contrastive log-ratio upper bound (vCLUB)~\cite{cheng2020club} is used to compute the upper bound of mutual information (MI) for irrelevant information of the $\textbf{e}$, decreasing the correlation among different clusters' representation:
\begin{equation}
\label{eq:mi}
\begin{split}
\mathcal{\hat{I}}(\textbf{e}_i, \textbf{e}_j)
=\frac{1}{\mathcal{N}^2}\sum_{\mathcal{M}=1}^\mathcal{N}\sum_{\mathcal{J}=1}^\mathcal{N}[\log f_\psi({\textbf{e}_i}_\mathcal{M}|{\textbf{e}_j}_\mathcal{M})
\\ -{\log f_\psi({{\textbf{e}_j}}_\mathcal{J}}|{\textbf{e}_i}_\mathcal{M})],
\end{split}
\end{equation}
where $\{\textbf{e}_i, \textbf{e}_j\} \in \textbf{e}$, $\mathcal{N}$ represents the samples from $\textbf{e}_i$ and $\textbf{e}_j$. $f_{\psi}(\textbf{e}_i|\textbf{e}_j)$ is a variational distribution with parameter $\psi$ to approximate $f(\textbf{e}_i|\textbf{e}_j)$. 
$\mathcal{\hat{I}}$ is the unbiased estimator for vCLUB with samples $\{{\textbf{e}_i}_\mathcal{M}, {\textbf{e}_j}_\mathcal{J}$.
The indexes $\mathcal{M}$ and $\mathcal{J}$ are the samples of $\textbf{e}_i$ and $\textbf{e}_j$. By minimizing Eq.~\ref{eq:mi}, we can decrease the correlation among
accent features $\textbf{e}_i$ and speech features $\textbf{e}_j$.
Finally, the central features of each cluster in $\mathcal{P}'$ are concatenated to form the codebook $\mathcal{E}_{codebook}$ columns in Figure~\ref{model}B.3. In summary, SECodec constructs a compressive codebook from complex speech feature graphs in an unsupervised and disentangled manner.

\subsection{Quantization via Node Game-based 2D SE Minimization}
The quantization of previous RVQ-based speech codec methods is to compare the input vectors with the vectors in the codebook and extract the indexes of the most similar vectors~\cite{defossez2022high,zeghidour2021soundstream}.
However, the comparison is performed by simply calculating the Euclidean distance between vectors. In high-dimensional space, the Euclidean distance tends to be uniformly distributed, which can distort the results and affect the quality of the quantized tokens. Consequently, the tokens obtained after quantization may lack sufficient real information when the size of the codebook is controlled within a reasonable range. To address this issue, SECodec views quantization as a process where graph nodes dynamically categorize subgraphs~\cite{zeng2024effective,yang2024incremental,sun2024lsenet} as shown in Figure ~\ref{model}C.1, treating new input features as added nodes in Figure ~\ref{model}C.2. In this approach, added nodes iteratively select the appropriate clusters through a structural entropy heuristic function, the selected clusters then correspond to the speech tokens as illustrated in Figure~\ref{model}C.3. Specifically, given a speech feature graph and its corresponding set of codebooks $E = [\textbf{e}_1,...,\textbf{e}_i,...,\textbf{e}_K]$. The added speech feature node $x$ selects the current codebook $\textbf{e}_i$, changing the codebook set to
$E' = [\textbf{e}_1,...,\textbf{e}_i',...,\textbf{e}_K,\{x\}](\textbf{e}_i=\textbf{e}'_i\cup\{x\})$. At this point,
the change of the graph’s 2D SE is formalized as:
\begin{equation}
\label{eq:quan}
\begin{split}
\Delta_{select}(x,\textbf{e}_i) = \mathcal{H}^{\mathcal{T}}(G)-\mathcal{H}^{\mathcal{T'}}(G)\\
=\sum_{n=1}^{|E|}H^{(2)}(\textbf{e}_n)-\sum_{n=1}^{|E'|}H^{(2)}(\textbf{e}'_n)\\
=-\frac{g_{\textbf{e}_i}}{\mathcal{V}_G}log\frac{\mathcal{V}_{\textbf{e}_i}}{\mathcal{V}_G}+\frac{g_{\textbf{e}'_i}}{\mathcal{V}_G}log\frac{\mathcal{V}_{\textbf{e}'_i}}{\mathcal{V}_G}\\-\frac{g_{\textbf{e}'_i}}{\mathcal{V}_G}log\frac{\mathcal{V}_{\textbf{e}'_i}}{\mathcal{V}_{\textbf{e}_i}}-\frac{d_x}{\mathcal{V}_G}log\frac{\mathcal{V}_{G}}{\mathcal{V}_{\textbf{e}_i}},
\end{split}
\end{equation}
where $\Delta_{select}(x,\textbf{e}_i)$ represents the change of the 2D SE when
node $x$ selects cluster $\textbf{e}_i$, and $\mathcal{T'}$ denotes the encoding tree corresponding to the codebooks set $\mathcal{E'}$. $\mathcal{H}^{\mathcal{T'}}(G)$ and $\mathcal{H}^{\mathcal{T}}(G)$ represent the 2D SE of the graph under $E$ and $E'$, respectively. $\mathcal{V}_G$, $\mathcal{V}_{\textbf{e}_i}$, and $\mathcal{V}_{\textbf{e}'_i}$
denote the volumes of the graph, cluster $\textbf{e}_i$, and cluster $\textbf{e}'_i$, respectively. $g_{\textbf{e}_i}$ and $g_{\textbf{e}'_i}$ correspond to the total cut edge weights of $\textbf{e}_i$ and $\textbf{e}'_i$, respectively. The added node only selects and joins
the codebooks cluster with the smallest change value of 2D SE, which is formalized as:
\begin{equation}
\label{eq:quan_min}
\begin{split}
t = Min(\Delta_{select}(x,\textbf{e}_i)),
\end{split}
\end{equation}
where $t$ represents the index of the target cluster, and the $Min$ operation finds the codebooks index corresponding to the
smallest 2D SE change value. 
\begin{algorithm}[!t]

\caption{Codebook construction via hierarchical and disentangled 2D SE
minimization.}\label{alg:alg1}
\label{algorithm:codebook}
\begin{algorithmic}
\STATE  \hspace{0.2cm} {\textbf{Input}}: Speech feature graph $G = (V,E,W)$, sub-graph size $n$.
\STATE  \hspace{0.2cm} {\textbf{Output}}: $\mathcal{E}_{codebook}= [\textbf{e}_1,...,\textbf{e}_K]$;
\STATE  \hspace{0cm} 1\hspace{0.2cm}$\mathcal{P} \leftarrow (x|x \in V)$
\STATE  \hspace{0cm} 2\hspace{0.2cm}\textbf{while} \textit{True} \textbf{do}
\STATE  \hspace{0cm} 3\hspace{0.2cm}\hspace{0.5cm}$\{\mathcal{P}_s\} \leftarrow$consecutively remove the first,
\STATE  \hspace{0cm} 4\hspace{0.2cm}\hspace{0.5cm}$min$$(n$,size of the remaining part of $\mathcal{P})$ clusters,
\STATE  \hspace{0cm} 5\hspace{0.2cm}\hspace{0.5cm}from $\mathcal{P}$ that form a set $\mathcal{P}_s$;
\STATE  \hspace{0cm} 6\hspace{0.2cm}\hspace{0.5cm}\textbf{for} $\mathcal{P} \in \{\mathcal{P}_s\}$ \textbf{do}
\STATE  \hspace{0cm} 7\hspace{0.2cm}\hspace{0.5cm}\hspace{0.5cm}$V' \leftarrow$ combine all the clusters in $\mathcal{P}_s$;
\STATE  \hspace{0cm} 8\hspace{0.2cm}\hspace{0.5cm}\hspace{0.5cm}$E' \leftarrow \{ e \in E$, both endpoints of $e \in V'\}$;
\STATE  \hspace{0cm} 9\hspace{0.2cm}\hspace{0.5cm}\hspace{0.5cm}$G' \leftarrow (V',E')$;
\STATE  \hspace{0cm}10\hspace{0.2cm}\hspace{0.5cm}\hspace{0.5cm}$\mathcal{T}' \leftarrow$ add a root tree node $\lambda$;
\STATE  \hspace{0cm}11\hspace{0.2cm}\hspace{0.5cm}\hspace{0.5cm}\textbf{for} \textit{cluster} $\mathcal{C} \in \mathcal{P}_s$ \textbf{do}
\STATE  \hspace{0cm}12\hspace{0.2cm}\hspace{0.5cm}\hspace{0.5cm}\hspace{0.3cm}Add a tree node $\alpha$ to $\mathcal{T}'$, $\alpha^-=\lambda, \mathcal{T}_{\alpha}=\mathcal{C}$;
\STATE  \hspace{0cm}13\hspace{0.2cm}\hspace{0.5cm}\hspace{0.5cm}\hspace{0.3cm}\textbf{for} \textit{speech feature} $x \in \mathcal{C}$ \textbf{do}
\STATE 
\hspace{0cm}14\hspace{0.2cm}\hspace{0.5cm}\hspace{0.5cm}\hspace{0.3cm}\hspace{0.3cm}Add a tree node $\gamma$ to $\mathcal{T}'$;
\STATE 
\hspace{0cm}15\hspace{0.2cm}\hspace{0.5cm}\hspace{0.5cm}\hspace{0.3cm}\hspace{0.3cm}$\gamma^-=\alpha, T_{\sigma}=\{x\}$;
\STATE
\hspace{0cm}16\hspace{0.2cm}\hspace{0.5cm}\hspace{0.5cm}$\mathcal{P}'\leftarrow$ run vanilla 2D SE minimization;
\STATE
\hspace{0cm}17\hspace{0.2cm}\hspace{0.5cm}\hspace{0.5cm}Append $\mathcal{P}'$ to $\mathcal{P}$;
\STATE  \hspace{0cm}18\hspace{0.2cm}\hspace{0.5cm}\textbf{if} $|\{V'\}|=1$ \textbf{then}
\STATE  \hspace{0cm}19\hspace{0.2cm}\hspace{0.5cm}\hspace{0.5cm}Break;
\STATE  \hspace{0cm}20\hspace{0.2cm}\hspace{0.5cm}\textbf{if} $\mathcal{P}$ \textit{is the same as at the end of last iteration} \textbf{then}
\STATE  \hspace{0cm}21\hspace{0.2cm}\hspace{0.5cm}\hspace{0.5cm} $n \leftarrow 2n$;
\STATE  \hspace{0cm}22\hspace{0.2cm}$\mathcal{P} = \{\mathcal{X}_1,\mathcal{X}_2,...,\mathcal{X}_K\}$
\STATE  \hspace{0cm}23\hspace{0.2cm}$\textbf{e}\leftarrow$[ ]
\STATE  \hspace{0cm}24\hspace{0.2cm}\textbf{for} $\mathcal{X}_i \in \mathcal{P}$ \textbf{do}
\STATE  \hspace{0cm}25\hspace{0.2cm}\hspace{0.5cm}$\textbf{e}_i \leftarrow 0$, count $\leftarrow 0$;
\STATE  \hspace{0cm}26\hspace{0.2cm}\hspace{0.5cm}\textbf{for} $x \in \mathcal{X}_i$ \textbf{do}
\STATE  \hspace{0cm}27\hspace{0.2cm}\hspace{0.5cm}\hspace{0.5cm}$\textbf{e}_i \leftarrow \textbf{e}_i+x$, count $\leftarrow$ count $+1$;
\STATE 
\hspace{0cm}28\hspace{0.2cm}\hspace{0.5cm}Append $\textbf{e}_i$ to $\textbf{e}$;
\STATE 
\hspace{0cm}29\hspace{0.2cm}$\textbf{e}=[\textbf{e}_1,...,\textbf{e}_K]$
\STATE 
\hspace{0cm}30\hspace{0.2cm}Minimize the mutual information in $\textbf{e}$ via Eq.~\ref{eq:mi} 
\STATE 
\hspace{0cm}31\hspace{0.2cm} $\mathcal{E}_{codebook}= [\textbf{e}_1,...,\textbf{e}_K]$
\STATE 
\hspace{0cm}32\hspace{0.2cm}\textbf{return} $\mathcal{E}_{codebook}$.

\end{algorithmic}
\label{alg1}

\end{algorithm}


\subsection{Training Objective}
In terms of training objective, we follow the setup of \cite{du2024funcodec}. The training objective consists of three components: reconstruction loss terms, adversarial loss terms, and RVQ commit losses. In the time domain, the L1 distance between the original speech and the reconstructed speech is minimized. In the frequency domain, both L1 and L2 distances are minimized across multiple Mel and magnitude spectra. For adversarial losses, SECodec incorporates several discriminators, including a multi-scale discriminator (MSD), a multi-period discriminator (MPD), and a multi-scale STFT-based (MSTFTD) discriminator.

\subsection{SESLM}
We build a structural entropy-based speech language model upon SECodec. Consisting of autoregressive and non-autoregressive models, it can hierarchically model information in
speech. Compared to VALL-E~\cite{wang2023neural}, we include the input of speech tokens in the autoregressive model part. We believe that the speech tokens extracted by SECodec are richer in speech information, which benefits speech language model training. The model learns to perform conditional generation of
a neural code sequence, denoted as $\mathcal{O}$,  based on two input prompts: textual prompt $u$ and acoustic prompt $\mathcal{S}$.
The training objective is formulated as:
\begin{equation}
\begin{split}
\mathcal{L}_{AR}=-\sum_{t=1}^{N}logP(\mathcal{O}_{t,1}|u, \mathcal{S}, \mathcal{O}_{\textless t,1}; \theta_{AR}),\\
\mathcal{L}_{NAR}=-\sum_{l=2}^{8}logP(\mathcal{O}_{:,l}|u, \mathcal{S}, \mathcal{O}_{:,\textless l}; \theta_{NAR}),
\end{split}
\end{equation}
where $\mathcal{O}_{\textless t,1}=[\mathcal{O}_{1,1},\ldots,\mathcal{O}_{t-1,1}]$, while $\theta_{AR}$ represents the AR Transformer model parameters. $\theta_{NAR}$ represents the NAR model parameters, while $\mathcal{O}_{:,l}$ denotes the entire sequence of $\mathcal{O}_{t,l}$ for the $l$th layer, and $\mathcal{O}_{:,\textless l}=[\mathcal{O}_{:,1},\ldots,\mathcal{O}_{:,l-1}]$ for $l=2,\ldots,8$. Note that the AR model in SESLM is conditioned on the
concatenated embeddings of both the acoustic and textual prompts. This formulation differs from that of VALL-E, where the AR model is only conditioned on the textual prompt and the past acoustic history. 
We validate the effectiveness of the structural entropy-based speech language model on the zero-shot TTS task. During inference, text input is converted to a phoneme sequence and the speech prompt to speech tokens. These are concatenated to form the prompts for both the AR and NAR models. The tokens generated by the AR and NAR models are then concatenated to construct the speech token matrix. Finally, the SECodec decoder is used to generate the waveform conditioned on the complete token matrix.

\begin{table*}[htb!]
\vspace{-1.0em}
\centering
\setlength{\tabcolsep}{1mm}{
    \begin{tabular}{l|c|cccccc}
    \hline

    Methods  &  $K$ in codebook   & {\textbf{WER $\downarrow$}} & {\textbf{MCD $\downarrow$}} & {\textbf{RMSE $\downarrow$}} & {\textbf{ViSQOL $\uparrow$}} & {\textbf{MUSHRA $\uparrow$}}& \textbf{Param}\\
    \hline
    Ground Truth &  -   & 4.58 & -  & - & -  & 91.46 & - \\
     EnCodec &  468  & 8.87 &  7.21 & 36.32 & 4.03  & 71.32 & 19.19M \\
      EnCodec &  1024  & 5.01 & 5.89  & 32.59 & 4.37  & 79.86 & 23.76M\\
      EnCodec &  2048  & 4.98 & 6.79  & 35.12 & 4.11  & 75.11 & 32.18M\\
      EnCodec &  4096  & 91.23 & 10.34  & 49.72 & -  & - & 82.71M\\
   SpeechTokenizer &  1024 & 5.04 & 6.23 & 34.84 & 4.30 & 90.55 & 120M\\
   \textbf{SECodec (Nodes:10000, Edges:\textgreater0.2)} &  468 &\textbf{4.63}& \textbf{5.12} & \textbf{31.97} & \textbf{4.40} & \textbf{90.79} & \textbf{19.19M}\\    
     \hline
   \multicolumn{5}{l}{\textit{Nodes Ablations} (Edges:\textgreater0.2)}\\
   SECodec (Nodes:100000) &  1212 &4.91& 6.37 & 32.79 & 4.31 & 88.21 &25.31M\\ 
   SECodec (Nodes:50000)&  768 &4.82& 5.57 & 32.45 & 4.37 & 90.12 &22.27M\\ 
   \hline
   \multicolumn{5}{l}{\textit{Edges Ablations} (Nodes:10000)}\\
   SECodec (Edges:\textgreater 0.5)&  1212&4.87& 6.19 & 31.81 & 4.29 & 86.33 &25.31M\\ 
   \hline
   \multicolumn{5}{l}{\textit{Methods Ablations}}\\
   SECodec \emph{w/o} SE Codebook&  468&4.89& 6.03& 32.37 & 4.32& 89.35 & 19.19M\\ 
   SECodec \emph{w/o} SE Quantization&  468&4.71& 5.67 & 33.89 & 4.39 & 87.62 & 19.19M\\ 
   \hline
    \end{tabular}
    
     \caption{Comparison of WER, MCD, RMSE, ViSQOL and MUSHRA of speech reconstruction on the LibriSpeech datasets.}
     \label{tab:results_SR}
    }
    \vspace{-1.0em}
\end{table*}

\section{Experiment
}
\subsection{Experimental Setup
}

\subsubsection{Data}

For SECodec training, we use the LibriSpeech~\cite{panayotov2015librispeech} dataset. At each training iteration, a 3.2 second segment is randomly cropped from the speech samples. For zero-shot TTS, we train AR and NAR models on the English subset of the Multilingual LibriSpeech dataset~\cite{pratap2020mls}, which contains 44,000 hours of transcribed speech data derived from LibriVox audiobooks. We select speech samples with durations ranging from 3 to 14 seconds for the training data. All speech data is sampled at a rate of 16 kHz.

\subsubsection{Model}
SEcodec is built on the framework of RVQ-GANs, following the same pattern as Funcodec~\cite{du2024funcodec}. SEcodec uses the convolutional-based encoder-decoder network from EnCodec, which performs temporal downscaling with a chosen striding factor. For zero-shot TTS experiments, AR model and NAR model are both 12-layer Transformer
decoders with 16 attention heads, an attention dimension of 1024 and the FFN dimension of 4096.

\subsubsection{Training}
During the training stage, we randomly clip a continuous segment
of 3.2 seconds from an utterance, which is considered as a training
sample. Before being fed into the encoder, the segment undergoes root-mean-square (RMS) normalization. The reconstructed output is rescaled using inverse normalization to calculate losses. We train the models on single 3090Ti GPUs with
a total batch size of 16. Under the adversarial training framework, we update the codec
model 300,000 times. To prevent the discriminator from becoming
too dominant, we only update it when its loss exceeds that of the
codec model.

\subsubsection{Baselines}
For SECodec, we consider two baselines: 1) EnCodec~\cite{defossez2022high}, and 2) SpeechTokenizer~\cite{zhang2023speechtokenizer}, a state-of-the-art speech tokenizer for speech language models. For SESLM, we consider two baselines: 1) VALL-E~\cite{wang2023neural}, and 2) ULSM~\cite{zhang2023speechtokenizer}.

\subsection{Metrics}

For speech reconstruction evaluation, we randomly sampled 300 speech samples from the LibriSpeech test set, considering both subjective and objective evaluation metrics. For objective metrics, we used mel-cepstrum distortion (MCD)~\cite{toda2007voice}, root mean square errors (RMSE)~\cite{luo2017emotional}, and ViSQOL~\cite{hines2015visqol} to assess speech quality. Additionally, we evaluated content accuracy through Word Error Rate (WER) by transcribing the speech using the Whisper en-medium model~\cite{radford2023robust}. For subjective metrics, we use MUSHRA following SpeechTokenizer~\cite{zhang2023speechtokenizer}.

For evaluating SESLM, we perform zero-shot text-to-speech assessments using the VCTK~\cite{veaux2016superseded} dataset, which includes recordings from 108 speakers with no overlap between the training data and the VCTK dataset. For each speaker, we randomly select a 3-second utterance as the prompt and use the text from a different utterance as the input. Objective metrics include evaluating speaker similarity and Word Error Rate (WER). To assess speaker similarity, we utilize the WavLM-TDCNN~\cite{chen2022wavlm} speaker embedding model to measure the similarity. We report the similarity with respect to the resynthesized audio context by its vocoder (SIM-r) and the similarity against the original audio context (SIM-o). As subjective metrics, we utilize Mean Opinion Score (MOS) for evaluating subjective audio quality (QMOS) and Similarity MOS (SMOS) for assessing subjective audio similarity. 

\begin{table*}[htb!]
\vspace{-1.0em}    
   
\centering
\setlength{\tabcolsep}{2mm}{
    \begin{tabular}{lc|ccccc}
    \hline

    Methods &Codec   & {\textbf{WER $\downarrow$}} & {\textbf{SIM-o $\uparrow$}} & {\textbf{SIM-r $\uparrow$}} & {\textbf{QMOS $\uparrow$}} & {\textbf{SMOS $\uparrow$}}\\
    \hline
    Ground Truth & - & 1.92 & 0.698  & n/a & 3.89($\pm0.18$)  & 3.92($\pm0.16$) \\
     SECodec resynthesis & SECodec  &2.02 & 0.682  & n/a & 3.82($\pm0.12$)  & 3.88($\pm0.11$) \\
   VALL-E & EnCodec& 7.09 & 0.501 & 0.412 & 3.08($\pm0.10$) & 3.31($\pm0.11$)\\
   USLM & SpeechTokenizer&5.79& 0.602 & 0.587 & 3.50($\pm0.13$) & 3.41($\pm0.12$)\\
   \textbf{SESLM (Nodes:10000, Edges:\textgreater0.2)} & \textbf{SECodec}&\textbf{4.97}& \textbf{0.634} & \textbf{0.611} & \textbf{3.63($\pm0.12$)} & \textbf{3.45($\pm0.11$)}\\
    \hline
   \multicolumn{5}{l}{\textit{Nodes Ablations} (Edges:\textgreater0.2)}\\
  SESLM (Nodes:100000) & SECodec&5.51& 0.612 & 0.593 & 3.56($\pm0.11$) & 3.37($\pm0.13$)\\ 
   \hline
    \multicolumn{5}{l}{\textit{Edges Ablations} (Nodes:10000)}\\
   SESLM (Edges:\textgreater 0.5) & SECodec&5.48& 0.602 & 0.597 & 3.46($\pm0.10$) & 3.33($\pm0.11$)\\  
    
     \hline
    \end{tabular}
    \caption{Comparison of WER, SIM-o, SIM-r, QMOS and SMOS of the proposed model and baselines on VCTK datasets.}
     \label{tab:results_tts}
    }
\vspace{-1.0em}
\end{table*}

\subsection{Main Results}

\subsubsection{Speech Reconstruction}
Table~\ref{tab:results_SR} provides a detailed summary of the speech reconstruction experiment results. SECodec outperforms both SpeechTokenizer and EnCodec by achieving a lower WER, highlighting its superior capability to retain content information. Furthermore, SECodec surpasses SpeechTokenizer and EnCodec in both MCD, RMSE, VISQOL and MUSHRA scores, underscoring its enhanced proficiency in producing high-quality speech.

\subsubsection{Zero-shot TTS}
Table~\ref{tab:results_tts} illustrates that our SESLM achieves a lower Word Error Rate (WER) compared to USLM and VALL-E. This finding underscores SECodec's capability to enhance the precision of content information modeling. Moreover, SESLM exhibits superior speaker similarity, suggesting that the modeled speech's structural information more effectively facilitates the extraction of paralinguistic features.

\subsubsection{Ablation Study}

Furthermore, we conducted an ablation study to analyze the performance effects of different components in SECodec, with the results presented in Table~\ref{tab:results_SR}
. The findings show that when SE quantization is removed, the model still performs well and outperforms most of the baseline models, but there is a noticeable decrease in speech quality. Additionally, when the SE codebook component is excluded, the results are significantly poorer, highlighting its critical importance to the overall performance.

\begin{figure*}[!t]

\centering
\includegraphics[width=6in]{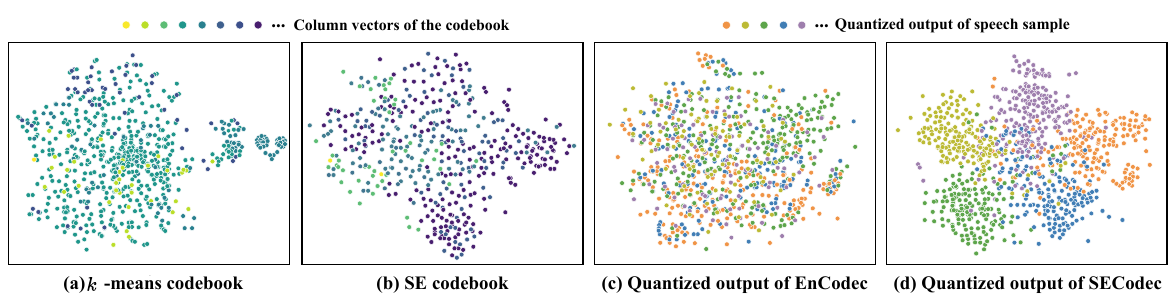}
\caption{Codebook and quantized output visualization.}
\label{figure_codebook}

\end{figure*}

\begin{figure*}[!t]
\vspace{-1.0em}
\centering
\includegraphics[width=6.1in]{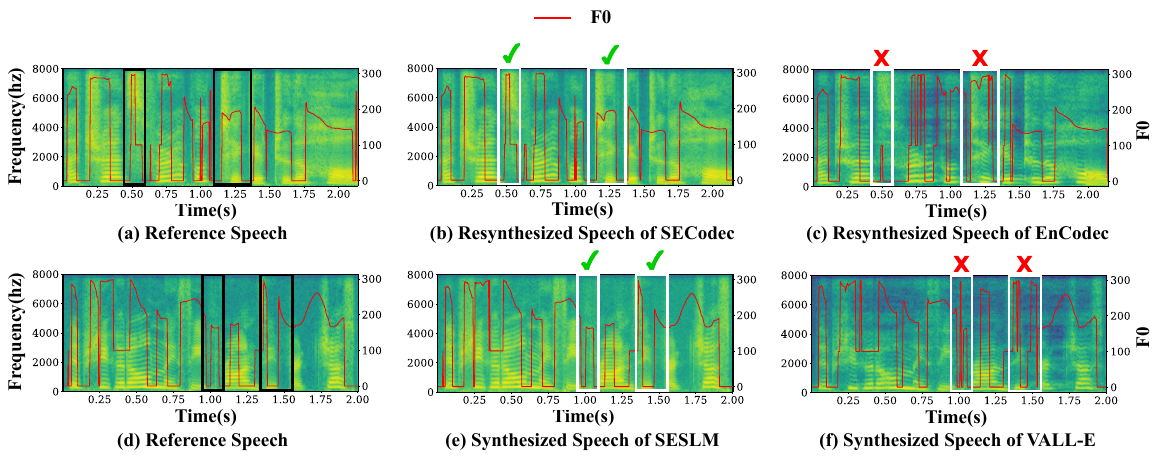}
\caption{Effectiveness of SE quantization. The black box indicates the structural details in the reference speech, the white box with a green checkmark indicates the part that our method correctly predicts, and the white box with a red cross indicates the part that the baseline method incorrectly predicts.}
\label{figure_quantization}
\vspace{-1.0em}
\end{figure*}

\subsection{Analysis}
\subsubsection{Choice of Nodes and Edges for SE}

We first analyzed the impact of different codebook sizes on the performance of EnCodec. It is evident that selecting an appropriate codebook size is crucial; a small codebook size (e.g., 468) leads to performance degradation, while an excessively large codebook (e.g., 4096) results in the model's failure to converge. Furthermore, we examined the influence of the number of nodes and edges in the speech feature graph during the SE minimization process. Table~\ref{tab:results_SR} presents the impact of varying the number of nodes and edges on the performance of SECodec. It can be observed that different numbers of nodes and edges, constructed with different similarity thresholds, lead to varying codebook sizes, which in turn affect the results. The best performance in speech reconstruction was achieved with 10,000 nodes and an edge threshold of 0.2. A similar conclusion is drawn from the ablation experiments on SESLM, as shown in Table~\ref{tab:results_tts}. On the other hand, SECodec achieves the best performance while also having the smallest model parameter size, further highlighting the importance of automatically selecting the appropriate codebook size.

\subsubsection{Codebook and Quantized Output Visualization}

To demonstrate the compressive and informative nature of the codebook learned by SECodec, we first visualized each column vector in the codebooks initialized by different methods using t-SNE. As illustrated in Figure~\ref{figure_codebook}(a) and Figure~\ref{figure_codebook}(b), the codebook initialized with k-means exhibits an uneven distribution in space, with cluster centers entangled and overlapping. In contrast, the codebook initialized with SECodec is more evenly distributed and discrete. To further illustrate the effectiveness of the codebook learned by SECodec, we visualized the quantized outputs of 1,000 speech samples. As shown in Figure~\ref{figure_codebook}(c) and Figure~\ref{figure_codebook}(d), the features of different speech samples obtained by SECodec are more disentangled and distinguishable compared to those obtained by EnCodec.

\subsubsection{Effectiveness of SE Quantization }

Figure~\ref{figure_quantization} displays the spectrogram and F0 (fundamental frequency) of both the reference speech and the resynthesized (synthesized) speech with identical content. It is evident that SECodec preserves the detailed features of the audio more effectively than EnCodec, particularly in maintaining the intricacies of the wave peaks at the fundamental frequency line (F0), as shown in Figure~\ref{figure_quantization}(a), Figure~\ref{figure_quantization}(b), and Figure~\ref{figure_quantization}(c). This observation is consistent in the speech synthesized by SESLM compared to VALL-E, as shown in Figure~\ref{figure_quantization}(d), Figure~\ref{figure_quantization}(e), and Figure~\ref{figure_quantization}(f), demonstrating that quantization via node game-based 2D SE minimization significantly reduces speech distortion caused by Euclidean distance-based quantization. Consequently, our method retains more detailed information in the synthesized speech.

\section{Conclusion}
Interacting with LLMs through speech has led to an increased demand for effective speech representation discretization. To address this, we propose SECodec, which can automatically determines the appropriate codebook size and integrates structural information into the quantization process.
Extensive experiments demonstrate that SECodec outperforms EnCodec in speech reconstruction. Furthermore, we developed a Structural Entropy-based Speech Language Model (SESLM) that leverages SECodec, yielding superior results in terms of generated speech content accuracy and quality. Additionally, the experiments show that SECodec is capable of learning a more compressive and discrete codebook and producing more informative speech tokens.

\section*{Acknowledgments}
This work was supported in part by the National Natural Science Foundation of China (Nos. U24A20334, 62376111, U23A2038 and 61972186), Yunnan provincial major scienceand technology special plan projects (Nos. 202302AD080003, 202402AG050007), Yunnan Provincial Key Researchand provincial major scienceand technology special plan projects (Nos. 202302AD080003, 202402AG050007), Yunnan Provincial Key Researchand Development Plan (Nos. 202302AD080003, 202402AG050007), Yunnan Provincial Key Research and Development Plan (Nos. 202303AP140008), Yunnan Province major basic research projects: 202401BC070021. The authors would like to thank anonymous reviewers for their comments.

\bibliography{aaai25}

\end{document}